\documentclass[10pt,conference]{IEEEtran}
\IEEEoverridecommandlockouts
\usepackage{cite}
\usepackage{amsmath,amssymb,amsfonts}
\usepackage{algorithmic}
\usepackage{graphicx}
\usepackage{textcomp}
\usepackage{xcolor}
\usepackage{booktabs}
\usepackage{tabularx}
\usepackage[hidelinks]{hyperref}
\usepackage{paralist}
\usepackage{xcolor}
\usepackage{balance}

\usepackage{subcaption}
\usepackage{tikz}
\usepackage{textcomp}
\usepackage{hyperref}

\newcommand\copyrighttext{%
  \footnotesize \textcopyright \the\year{} IEEE. Personal use of this material is permitted.  Permission from IEEE must be obtained for all other uses, in any current or future media, including reprinting/republishing this material for advertising or promotional purposes, creating new collective works, for resale or redistribution to servers or lists, or reuse of any copyrighted component of this work in other works. DOI: \url{https://doi.org/10.1109/SANER60148.2024.00048}}
\newcommand\copyrightnotice{%
\begin{tikzpicture}[remember picture,overlay]
\node[anchor=south,yshift=10pt] at (current page.south) {\fbox{\parbox{\dimexpr\textwidth-\fboxsep-\fboxrule\relax}{\copyrighttext}}};
\end{tikzpicture}%
}

\newcommand{\alg}{\textsc{ShRec}}
\def\BibTeX{{\rm B\kern-.05em{\sc i\kern-.025em b}\kern-.08em
    T\kern-.1667em\lower.7ex\hbox{E}\kern-.125emX}}

\begin{document}

\title{\alg: a SRE Behaviour Knowledge Graph Model for Shell Command Recommendations
\thanks{$^{\dag}$Corresponding author (andrea.tonon1@huawei-partners.com).}
}

\author{\IEEEauthorblockN{Andrea Tonon,$^{1,\dag}$ Bora Caglayan,$^1$ MingXue Wang,$^1$ Peng Hu,$^2$  Fei Shen,$^2$ Puchao Zhang$^1$}
\IEEEauthorblockA{\textit{$^1$Huawei Ireland Research Center}, Dublin, Ireland \\
\textit{$^2$Huawei Nanjing R\&D Center}, Nanjing, China}}

\maketitle

\copyrightnotice

\begin{abstract}
In IT system operations, shell commands are common command line tools used by site reliability engineers (SREs) for daily tasks, such as system configuration, package deployment, and performance optimization. The efficiency in their execution has a crucial business impact since shell commands very often aim to execute critical operations, such as the resolution of system faults. However, many shell commands involve long parameters that make them hard to remember and type. Additionally, the experience and knowledge of SREs using these commands for analysing or troubleshooting is almost always not preserved.
In this work, we propose \alg, a SRE behaviour knowledge graph model for shell command recommendations. We model the SRE shell behaviour knowledge as a knowledge graph and propose a strategy to directly extract such a knowledge from SRE historical shell operations. The knowledge graph is then used to provide shell command recommendations in real-time to improve the SRE operation efficiency. Our empirical study based on real shell commands executed in our company demonstrates that \alg\ can improve the SRE operation efficiency, allowing to share and re-utilize the SRE knowledge. 
\end{abstract}


\section{Introduction}
\begin{figure}[]
\centerline{\includegraphics[width=0.9\linewidth]{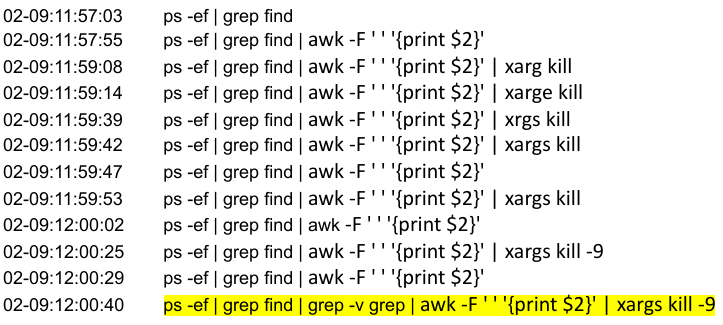}}
\caption{Example of errors found in real shell data before the execution of the correct command (highlighted in yellow).}
\label{fig:err}
\end{figure}

Site reliability engineers (SREs) every day perform a huge number of IT system operations, such as system configuration, package deployment, and performance optimization, through the execution of shell commands. 
The efficiency in executing such commands is of crucial importance in many business scenarios, since IT operations may represents key tasks required to solve critical system faults that may cause thousands of dollars in losses in large IT companies such as ours. Unfortunately, most shell commands require long and complex parameters, making them hard to remember and type~\cite{gandhi2020lightening}. For example, by checking shell commands executed by SREs of our company, we found that the command 
\begin{equation*}
\begin{split}
 &\textit{cat /opt/hw/app/common/business\_flume\_client/} \\
 &\textit{flume\_1.5/conf/properities.properties $|$ grep topics}
\end{split}
\end{equation*}
is frequently executed to just check the Flume service
configuration. In addition, we found that erroneous commands precede the execution of a correct shell command in many cases, in particular for less experienced SREs, as shown in Fig.~\ref{fig:err} where the execution of the correct command required more than 3 minutes and the execution of 12 commands. Thus, to improve the SRE operation efficiency and to share the SRE knowledge to all SREs are tasks with a significant business impact.

In such a direction, standard shell suggestion systems~\cite{fish,zsh} provide auto-complete functionalities considering historical commands.
However, they lack a knowledge model to preserve the SRE operational knowledge for different IT operations, such as where to find a particular configuration file, which sequence of commands is usually executed to check a given service status, etc. Thus, they do not allow to preserve nor re-utilize the SRE behaviour knowledge of shell operations that can be found in historical shell command data. Additionally, they almost always consider exact matches between commands and do not provide sequence based suggestions.

\begin{figure*}
        \centering
        \begin{subfigure}[b]{0.49\linewidth}
            \centering
            \includegraphics[width=\linewidth]{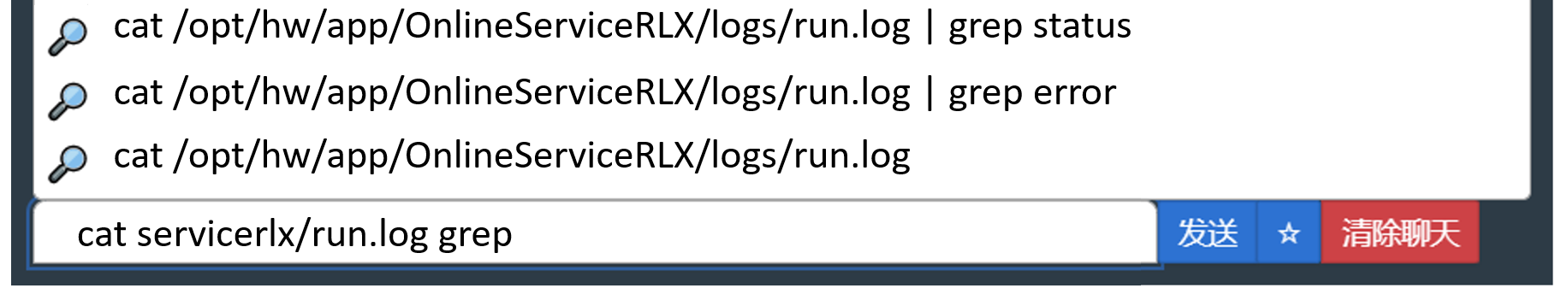}
    \end{subfigure}
        \hfill
        \begin{subfigure}[b]{0.49\linewidth}  
            \centering 
            \includegraphics[width=\linewidth]{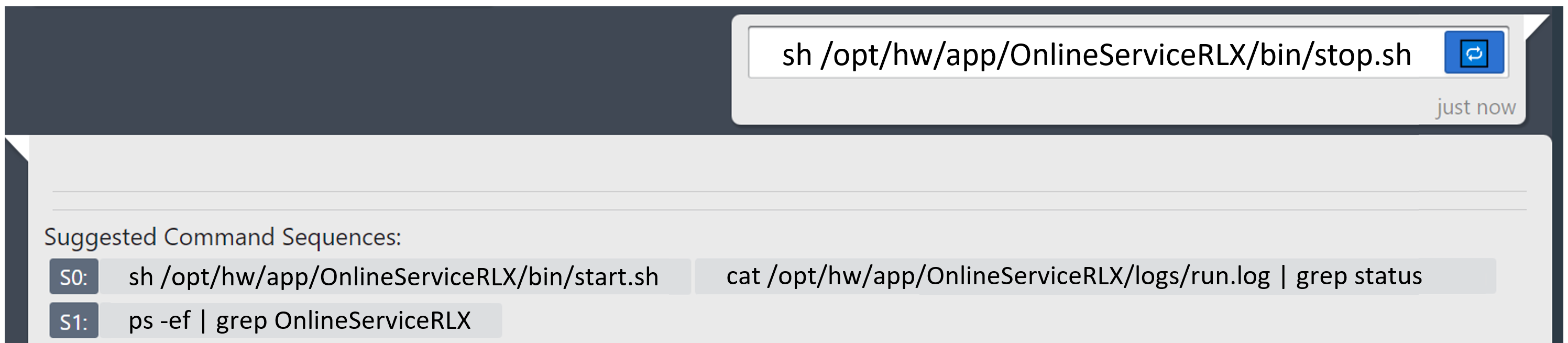}
    \end{subfigure}
    \caption{Example of command (left) and sequence (right) recommendations provided by \alg.} 
        \label{fig:recEXX}
    \end{figure*}

In this work, we introduce \alg, a novel method to model and recommend shell commands and sequences of shell commands for IT operations to improve the SRE operation efficiency. In particular, we first collect SRE historical shell data, representing shell commands executed by SREs to solve IT system operations in our company. From the historical shell data, we then automatically extract SRE behaviour knowledge than can be reused to provide recommendations. Let us note that the SRE experience in using such commands for analysing or troubleshooting is almost always not preserved. Through the shell command parsing and processing, SRE behaviour pattern mining, SRE behaviour pattern aggregation, and SRE intent definition processes, we are thus able to automatically extract and classify commands and sequences of commands representing useful and complex operations executed by SREs. (Note that the manual definition of such a knowledge, as done in other works~\cite{kurs2016nextflowworkbench} for different tasks, would require a lot of effort.) In addition, by automatically extracting SRE knowledge from historical data, our approach allows to directly learn additional relations between the involved entities, such as IPs, users, business scopes, etc., and their respective execution frequency. The extracted SRE behaviour knowledge is then modeled in a knowledge graph, i.e., the SRE behaviour knowledge graph, that allows to preserve and to represent the SRE knowledge and the relations between the involved entities. The SRE behaviour knowledge graph is then employed by our knowledge graph based recommender system to recommend commands and sequences of commands allowing to share the SRE knowledge to all the SREs improving their efficiency. 
Fig.~\ref{fig:recEXX} shows an example of command and sequence recommendations provided by \alg. 
For example, while the user is typing `\textit{cat servicerlx/run.log grep}', real-time command recommendations allow to directly obtain the whole path of the file and possible arguments for the `\textit{grep}' command, reducing the information that SREs have to memorize and the characters to type.
In addition, after the user executed a command, such as `\textit{sh /opt/hw/app/OnlineServiceRLX/bin/stop.sh}' to stop a service, the sequence recommendations provide sequences of shell commands that can continue such an operation. For example, the first recommended sequence allows to restart such a service and to check its log file, while the second one allows to check if the service has been correctly stopped, all operations executed very often after the first command. Thus, sequence recommendations allow to directly execute shell commands without the exigence of typing them.  

In this regard, our contributions are:
\begin{itemize}
    \item We introduce a method to extract SRE knowledge of shell operations from historical shell data. Our approach allows to automatically extract shell commands and sequences of shell commands representing complex operations executed by SREs and their relations with additional entities, such as IPs, users, files, intents, etc. Such experience is almost always not preserved nor re-used.  
    \item We design a SRE behaviour knowledge graph model to preserve the SRE operational knowledge learned from shell data. The knowledge model contains (sequences of) shell commands and their relations with further entities, allowing to re-utilize them to provide recommendations.
    \item We introduce a recommender algorithm that employs the knowledge graph model to recommend (sequences of) shell commands considering context information, such as IPs, users, business scopes, etc, to share the SRE knowledge learned from shell data to all the SREs.  
    \item We discuss the SRE operation efficiency improvement that \alg\ can provide considering statistics estimated on real data, showing the benefits that to preserve and to re-utilize SRE shell operational knowledge can provide.
\end{itemize}

\section{Related Works}
We now discuss the relation of our work to prior art on shell suggestion systems and sequential recommender systems employing pattern mining techniques or knowledge graphs.

Shell commands are widely used for accomplishing tasks such as network management and file manipulation. Given the large number of shell commands available and the long parameters that most of them require, the development of systems that provide auto-complete and recommendation functionalities~\cite{fish,zsh} had a fair success. However, such methods directly take into account the executed command history to provide auto-completions, almost always considering exact matches between the commands.
On the other hand, \cite{zhang2022shellfusion} proposed ShellFusion, an approach to automatically generates comprehensive answers for shell programming tasks considering shell knowledge mined from question/answer posts and public available tutorials. By considering such data resources, ShellFusion generates comprehensive answers for general shell tasks, while in our work we focus on specific shell commands executed by SREs of our company. Additionally, our method directly recommends shell commands that can be executed in real-time instead of providing comprehensive answers. Instead, other works~\cite{anoshin2020getting,hu2018loquat,kurs2016nextflowworkbench} describe systems to provide command recommendations for other tasks, such as SQL programming~\cite{anoshin2020getting}, but they consider manually defined commands~\cite{kurs2016nextflowworkbench} and/or do not provide sequence recommendations. Thus, with all these approaches, the SRE behaviour knowledge of shell operations cannot be preserved nor re-utilized.     

Data mining techniques have already been employed to extract key insights for conceptual modeling~\cite{fumagalli2022pattern,da2022discovery}. In such a direction, in this work, we consider the sequential pattern mining framework~\cite{agrawal1995mining}. Since the introduction, several algorithms have been proposed for this task~\cite{ayres2002sequential,pei2004mining,tonon2019permutation}. In particular, sequential patterns have been successfully applied in recommender systems~\cite{yap2012effective} to model the sequential nature of the data, but they are usually employed to provide next-item recommendations of new unseen elements instead of to extract complex operations frequently executed, as done in this work. 

Further works~\cite{huang2019explainable,li2023preference} consider knowledge graphs as additional resources to improve performance and explainability of recommender systems. On the contrary, we model the knowledge graph to preserve the SRE knowledge learned from the data, and use it to provide recommendations.

To the best of our knowledge, this is the first work to model a knowledge graph to represent and preserve SRE knowledge automatically learned from shell data to provide recommendations of shell commands considering context information.

\section{Preliminaries}
We now provide concepts and definitions used in the paper.

\subsection{Sequential Pattern Mining}
\label{sec:fsp}
Let $\mathcal{I} = \{i_1, i_2, \dots, i_h\}$ be a finite \emph{ground set} of elements called \emph{items}. A \emph{sequential pattern}, or \emph{sequence} $s = \langle i_{j_1}, i_{j_2}, \dots, i_{j_\ell} \rangle$ is a \emph{finite ordered list} of $\ell$ items. (Let us note that in other works, a sequential pattern is defined as a finite ordered list of \emph{sets of items}, while in this work we provide a simplified version of such a definition that better fits for our scenario.) The \emph{length} $|s|$ of $s$ is the number of items in $s$. A sequence  $a = \langle a_{1}, a_{2}, \dots, a_{|a|} \rangle$ is a \emph{sub-sequence} of another sequence $b = \langle b_{1}, b_{2}, \dots, b_{|b|} \rangle$ with respect to (w.r.t.) a \emph{maximum gap} $g \in \mathbb{N}^+$, denoted by $a \sqsubseteq_g b$, if and only if there exist integers $1 \leq r_1 < r_2 < \dots < r_k \leq |b|$ such that $a_1 = b_{r_1}, a_2 = b_{r_2}, \dots, a_k = b_{r_k}$, and, for each pair of consecutive items $a_j, a_{j+1} \in a$, $r_{j+1} - r_j \leq g$, with $j \in \{1, |a| -1\}$. A \emph{dataset} $\mathcal{D}$ is a finite bag of $|\mathcal{D}|$ transactions, $\mathcal{D} = \{\tau_1, \tau_2,\dots, \tau_{\mathcal{D}}\}$, where each transaction $\tau \in \mathcal{D}$ is a sequential pattern with items from the ground set $\mathcal{I}$. A sequence $s$ \emph{belongs} to a transaction $\tau \in \mathcal{D}$ w.r.t. a maximum gap $g \in \mathbb{N}^+$ if and only if $s \sqsubseteq_g \tau$. For any sequence $s$, the \emph{support} $supp_{\mathcal{D}}(s,g)$ of $s$ in $\mathcal{D}$ w.r.t. $g$ is the number of transactions in $\mathcal{D}$ to which $s$ belongs w.r.t. $g$, i.e.,
$supp_{\mathcal{D}}(s,g) = |\{\tau \in \mathcal{D} : s \sqsubseteq_g \tau\}|$.
Finally, the \emph{frequency} $f_{\mathcal{D}}(s,g)$ of $s$ in $\mathcal{D}$ w.r.t. $g$ is the fraction of transactions in $\mathcal{D}$ to which $s$ belongs w.r.t. $g$, i.e.,
$f_{\mathcal{D}}(s,g) = supp_{\mathcal{D}}(s,g) / |\mathcal{D}|$.
Let $\mathbb{S}$ denote the set of all the possible sequences built with items from $\mathcal{I}$. Given a dataset $\mathcal{D}$, a \emph{minimum frequency threshold} $\theta \in (0,1]$, and a maximum gap $g \in \mathbb{N}^+$, the \emph{sequential pattern mining} task requires to output the set $FSP(\mathcal{D},\theta,g)$ of all sequences from $\mathbb{S}$ whose frequencies in $\mathcal{D}$ w.r.t. $g$ are at least $\theta$, and their frequencies, i.e., 
$FSP(\mathcal{D},\theta,g) = \{(s, f_\mathcal{D}(s,g)) : s \in \mathbb{S}, f_\mathcal{D}(s,g) \geq \theta\}$.

\subsection{Knowledge Graph}
A \emph{knowledge graph} is a graph-structured data model that captures semantic relationships between entities such as events, objects, or concepts. This information is usually stored in a graph database and visualized as a graph structure, prompting the term knowledge graph.
Since there is no single commonly accepted definition of a knowledge graph, we now discuss the definition that we consider in this work. In particular, we consider a knowledge graph as a graph in which the vertices represent entities and the edges represent relationships between the entities. Without loss of generality, we represent the relationships as undirected edges, which can be traversed in either directions. Each vertex is identified by a unique vertex ID and has a tag representing the type of entity. For example, in our knowledge graph, a tag is \textit{user}, with each vertex of such a type representing a different user. An edge, instead, represents a connection or a behaviour between two vertices, defined by the types of vertices it connects. For example, an edge connecting a vertex $A$ of type \textit{user} and a vertex $B$ of type \textit{cmd} represents that the user represented by $A$ executed the command represented by $B$. Both vertices and edges can have some properties, with same type vertices and same type edges sharing the same definition of properties, respectively. Accessing a vertex by considering its ID, it is possible to reach all vertices connected to it by a given edge type, eventually conditioning on vertices and edges' properties.   

\subsection{Historical Shell Command Data}
Shell programming is widely used to accomplish many tasks. In this work, we consider to collect the shell commands that every day the SREs of our company execute, obtaining a collection of \emph{SRE historical shell command data}, simply denoted as \emph{shell data}. For each executed command, we considered the following entities:
\begin{inparaitem}
    \item command: the executed shell command;
	\item scope: a high-level classification about the business scope, i.e., system or service, under which the command has been executed;
	\item timestamp: the timestamp in which the shell command has been executed;
	\item user: the user that executed the shell command.
\end{inparaitem}

Such shell commands are grouped into \emph{sessions}, i.e., sequences of commands executed inside $ssh$ sessions. Each session starts with a $ssh$ command that creates a connection with a given server, associated with an IP address, and contains all the commands executed in that server by a given user under a unique scope. The scope, instead, represents a high-level classification about the system/service under which the commands are executed. In particular, commands executed in different servers but under the same scope access the same file-system. Thus, recommendations make sense only under the same scope. Note that these choices are dictated by the system architecture from which we collected the data. However, \alg\ can also consider different system architectures. For example, in a system in which each server has its own file-system, it is possible to provide recommendations considering the IP address of the server, instead of considering the scope.        

\section{\alg: Method Overview}
\begin{figure*}[]
\centerline{\includegraphics[width=0.9\textwidth]{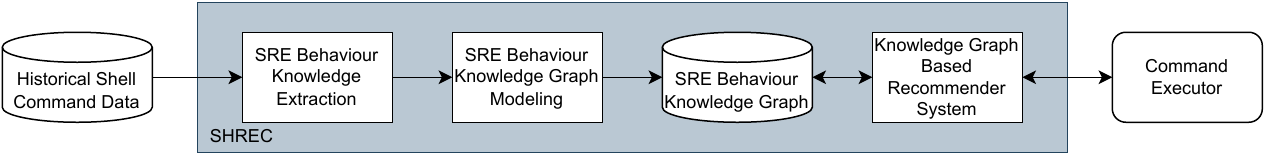}}
\caption{\alg\ overview.}
\label{fig:system}
\end{figure*} 
In this section, we introduce the overall architecture of \alg: a SRE behaviour knowledge graph model for \underline{SH}ell command \underline{REC}ommendations. Fig.~\ref{fig:system} shows a schema with all its components:
\begin{inparaenum}
\item historical shell command data: all the collected shell commands executed by SREs and their information;
\item SRE behaviour knowledge extraction: starting from the shell data, it extracts SRE knowledge that can be re-used and shared between SREs. It consists in shell command parsing and processing, SRE behaviour pattern mining, SRE behaviour pattern aggregation, and SRE intent definition;
\item SRE behaviour knowledge graph modeling: it models the extracted SRE knowledge in the SRE behaviour knowledge graph;
\item SRE behaviour knowledge graph: it stores entities and relations extracted during the SRE behaviour knowledge extraction;
\item knowledge graph based recommender system: after a recommendation request received from the command executor, it retrieves from the SRE behaviour knowledge graph all the information required to provide recommendations. Through our ranking procedure, it provides personalized recommendations based on the received request. It consists in command recommender and sequence recommender;  
\item command executor: an external application that allows the users to execute shell commands. It sends recommendation requests to the knowledge graph based recommender system and shows the received recommendations to the users.
\end{inparaenum}

In the next sections, we provide a description of its parts.

\section{SRE Behaviour Knowledge Extraction}
The aim of the SRE behaviour knowledge extraction is to extract SRE knowledge that can be re-used and shared between SREs. It consists in shell command parsing and processing, SRE behaviour pattern mining, SRE behaviour pattern aggregation, and SRE intent definition.
\subsection{Shell Command Parsing and Processing}
\label{sec:pre}
In this section, we describe how we processed the shell data. The idea is to prepare the data for the next phases, removing incorrect commands and parsing them to extract additional information, such as IP addresses from `\textit{ssh}' commands and accessed paths/files from commands associated with files (e.g., `\textit{cat}', `\textit{vi}', etc.). For such commands, we also aim to convert relative paths to absolute paths to reduce data sparsity. 

By following the execution flow of the commands, we parsed the shell commands contained in each session in the shell data. While following the execution flow of the commands inside a session, we also maintained the path of the user location by considering the executed `\textit{cd}' commands. We thus parsed each shell command, removing its not-crucial options and extracting its arguments. Whether it was a command associated with a file, we also converted the path used to access the file to an absolute path by considering the user location that we maintained following the execution flow. For example, the command  `\textit{cat logs/result.log}’ executed while the user is located in `\textit{/data}' becomes `\textit{cat /data/logs/result.log}'. In this phase, we also removed commands containing syntax errors (i.e., errors that can be detected while parsing the commands), such as wrong arguments and/or options, and not-crucial commands, such as sequences of `$cd$' commands that can be replaced by a single `$cd$' command considering the last location. Finally, we removed all the commands that appeared in a number of sessions $< min\_supp$, with $min\_supp \in \mathbb{N}$ a \emph{minimum support threshold}, with the idea that such rare commands are errors or not-useful commands.    

\subsection{SRE Behaviour Pattern Mining}
\label{sec:patMin}
In this section, we describe how we employed the sequential pattern mining framework to extract, from the processed data, sequences of commands frequently executed in the shell data.
In our scenario, each shell command is considered an item. Then, the temporal sequence of commands that composes a session represents a transaction, and the collection of all the sessions represents the input dataset. By employing any sequential pattern mining algorithm, it is thus possible to mine frequent sub-sequences of commands that have been executed in many different sessions and that represent useful operations that can be recommended over times. Note that in this scenario, the maximum gap constraint (see section~\ref{sec:fsp}) is useful for two reasons. Since commands that composed an operation are almost always executed close together, by considering a maximum gap between the commands, it is possible to avoid sequences that appear frequent due to some noise in the data and thus that do not represent useful operations. In addition, by reducing the research space, the mining phase can be speeded up of several orders of magnitude. In such a direction, based on the requirements, it is also possible to add a minimum and/or maximum sequence size constraint. Finally, after the mining, additional filters can be applied to the mined sequences.
For example, it is possible to remove sequences that have been executed only by a few users, since they may represent too specific operations, or only in a single day, since they may be associated with too rare events.
In addition, to reduce data repetition, it is possible to remove (consecutive) repetitions of commands inside the sequences and/or to remove sequences $a$ whether there exists at least a sequence $b$ such that $a \sqsubseteq_g b$ and  
$f_{\mathcal{D}}(b,g) \geq r \cdot f_{\mathcal{D}}(a,g)$,
with $r \in (0,1)$ a \emph{redundancy threshold}. The idea of this last filter is to remove sequences that are contained in other mined sequences if their frequencies are close enough, since the smallest sequences do not provide further information.
An example of mined sequence is
\begin{equation*}
\begin{split}
 &\textit{1) cat /opt/hw/app/OnlineServiceRLX/conf/app.properties} \\
 &\textit{2) sh /opt/hw/app/OnlineServiceRLX/bin/stop.sh} \\
 &\textit{3) sh /opt/hw/app/OnlineServiceRLX/bin/start.sh} \\
 &\textit{4) cat /opt/hw/app/OnlineServiceRLX/logs/run.log}.
\end{split}
\end{equation*}
It is composed by 4 shell commands, it represents that a user checked a configuration file of the OnlineServiceRLX service, restarted such a service, and finally checked its log file, and it has been executed in 8 sessions by 2 users in 2 days.

\subsection{SRE Behaviour Pattern Aggregation}
\label{sec:patAgg}
In this section, we describe how the sequences extracted as explained in the previous section can be aggregated to obtain more concise operation definitions.
In particular, let us note that there may be some mined sequences that represent almost the same operations, for example just differing for some command arguments. Since to maintain all such sequences is not useful, they can be used to manually define general sequences of commands, called \emph{macros}, that better summarize the operations that the similar sequences represent. Obviously, SREs can manually define macros also directly considering the shell data, but this would require to analyze a huge amount of data. Instead, we propose a clustering based method to group together similar frequent sequences to analyze them more efficiently. Starting from the mined sequences, we first computed the distance between every pair of sequences, obtaining a distance matrix. Given two sequences of commands $x = \langle x_1, x_2, \dots, x_{|x|} \rangle$ and $y = \langle y_1, y_2, \dots, y_{|y|} \rangle$, we computed their distance $dist(x,y)$ as

\footnotesize   
\begin{equation*}
 dist(x,y) = \frac{\sum_{i=1}^{\min{(|x|,|y|)}}[distJ(x_i, y_i)] + \max(|x|,|y|) - \min(|x|,|y|)}{ \max(|x|,|y|)}
\end{equation*}
\normalsize
where $distJ(x_i, y_i)$ is the Jaccard distance between the i-th command of $x$ and the i-th command of $y$ after that they are tokenized considering spaces and path components (e.g., \textit{`cat /data/logs/result.log'} = [\textit{cat, data, logs, result.log}]). In particular, the idea is to compute their distance considering the number of tokens that they share. (Since absolute paths can be very long, we found that to separate their components allows to represent their distance more accurately.) The factor $\max(|x|,|y|) - \min(|x|,|y|)$ takes into account the possible different length of $x$ and $y$, while the factor $\max(|x|,|y|)$ normalizes the distance. Then, we employed a clustering algorithm (e.g., k-means) to cluster similar sequences together, by using the distance matrix. (The optimal number of clusters can be computed, for example, using the silhouette coefficient.) Finally, by analyzing the clustered sequences, SREs can decide whether manually define macros that better summarize the operations that the clustered sequences represent, assigning to each macro an \emph{intent} that describes its aim, potentially with some parameters. For example,
the intent \textit{`restart\_service=Y'} can be used to define a sequence like the one showed in section~\ref{sec:patMin}. The intent can be used to retrieve the respective sequence from the knowledge graph for directly executing it.

\subsection{SRE Intent Definition}
\label{sec:intent}
In this section, we describe how we defined SRE intents for the shell commands, similarly to what we did in the previous section for the mined sequences. The idea is to define intents for the shell commands to obtain more concise and general commands to further simplify their execution. Thus, SREs do not have to memorize and explore long paths and complex parameters to execute shell commands. Based on the shell data, we defined 8 intents to represent the most frequent operations, i.e., `\textit{log\_analysis fileName}', `\textit{config\_analysis fileName}', `\textit{process\_analysis [processName]}', `\textit{crontab\_analysis [processName]}', `\textit{storage\_analysis}', `\textit{network\_analysis}', `\textit{execute\_script fileName}', and `\textit{code\_analysis fileName}', and for each of them we defined a set of rules to automatically classify the shell commands into intents. For example, for the `\textit{log\_analysis fileName}' intent, we considered all the commands associated with files, e.g., `\textit{cat}', `\textit{vi}', accessing files with an extension or a path associated with log files, e.g., `\textit{.log}', `\textit{.dat}', `\textit{/logdir/}', `\textit{/interface\_logs/}. By defining similar rules for all the intents, we can automatically classify shell commands into intents and use such intents to directly retrieve the respective commands from the knowledge graph.      
\begin{figure*}[h]
\centerline{\includegraphics[width=0.75\linewidth]{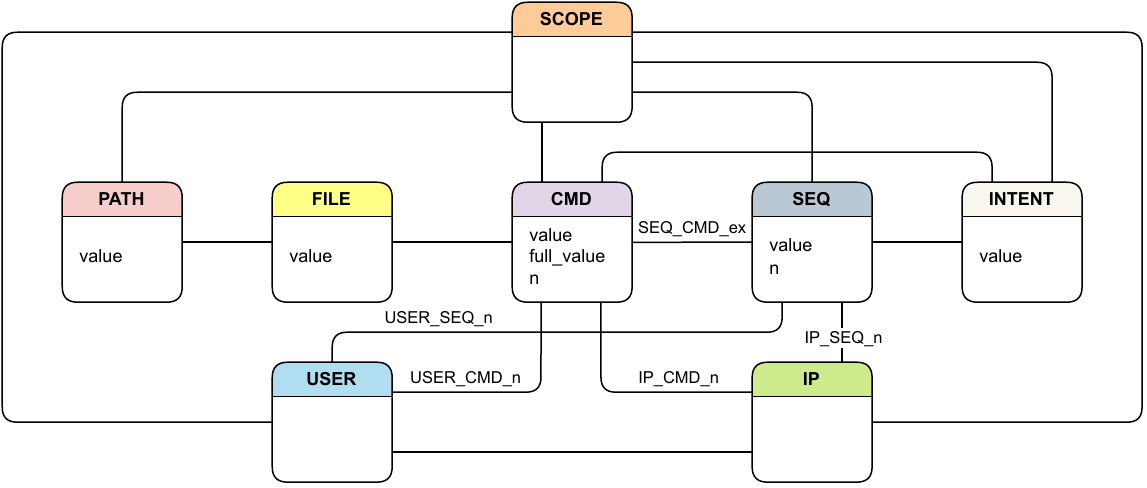}}
	\caption{Schema of the SRE behaviour knowledge graph.}
	\label{fig:kg}
\end{figure*} 
\section{SRE Behaviour Knowledge Graph Modeling}
\label{sec:kg}
In this section, we describe how we designed the knowledge graph to store and preserve the SRE knowledge extracted in the previous phases, representing the relations between the different entities. Fig.~\ref{fig:kg} reports the schema of the SRE behaviour knowledge graph. We considered 8 types of vertices:
\begin{itemize}
\item \textit{scope}: represents the scopes in the shell data. Each vertex is associated with a scope and its ID is such a scope;
\item \textit{user}: represents the users in the shell data. Each vertex is associated with a user and its ID is such a user's ID;
\item \textit{IP}: represents the IPs of the sessions in the shell data. Each vertex is associated with an IP and its ID is such an IP;
\item \textit{path}: represents the absolute paths extracted from the shell commands in the shell data. Each vertex is associated with a pair (scope,path) and has an ID based on it. The property \textit{path.value} is the absolute path represented by such a vertex (e.g., `\textit{/data/logs/}');
\item \textit{file}: represents the files accessed in the shell data by shell commands. Each vertex is associated with a triple (scope,path,file) and has an ID based on it. The property \textit{file.value} is the file name represented by such a vertex (e.g., `\textit{result.log}');
\item \textit{cmd}: represents the commands executed in the shell data. Each vertex is associated with a pair (scope,cmd) and has an ID based on it. The properties \textit{cmd.value} and \textit{cmd.full\_value} are, respectively, the type of shell command (e.g., `\textit{cat}') and the whole shell command (e.g., `\textit{cat /data/logs/result.log}') represented by such a vertex, while the property \textit{cmd.n} is the number of times the command has been executed in the shell data under the associated scope. For commands that directly execute a file (e.g., `\textit{./scripts/bin/startup.sh}'), \textit{cmd.value} = `\textit{execute}';
\item \textit{seq}: represents the sequences mined from the shell data and the macros defined by SREs. Each vertex is associated with a pair (scope,seq) and has an ID based on it. The property \textit{seq.value} is the whole sequence of commands represented by such a vertex, while the property \textit{seq.n} is the number of times the sequence has been executed in the shell data under the associated scope;
\item \textit{intent}: represents the intents of the macros defined by SREs and of the classified shell commands. Each vertex is associated with a pair (scope,intent) and has an ID based on it. The property \textit{intent.value} is the intent represented by such a vertex. 
\end{itemize}

The vertices are connected with edges as shown in Fig.~\ref{fig:kg}. Let us note that only commands that accessed files (e.g., `\textit{cat /data/logs/result.log}') are connected with the respective files, only commands contained in mined sequences are connected with the respective sequences, and only commands and sequences with an intent are connected with the respective intents. Additionally, five types of edges have a property:
\begin{itemize}
\item \textit{user\_cmd\_n}: a property of the edges connecting a \textit{user} vertex with a \textit{cmd} vertex. It represents how many times the user executed the command;
\item \textit{IP\_cmd\_n}: a property of the edges connecting a \textit{cmd} vertex with an \textit{IP} vertex. It represents how many times the command has been executed inside a session associated with the IP;
\item \textit{user\_seq\_n}: a property of the edges connecting a \textit{user} vertex with a \textit{seq} vertex. It represents how many times the user executed the sequence;
\item \textit{IP\_seq\_n}: a property of the edges connecting a \textit{seq} vertex with an \textit{IP} vertex. It represents how many times the sequence has been executed inside a session associated with the IP;
\item \textit{seq\_cmd\_ex}: a property of the edges connecting a \textit{seq} vertex with a \textit{cmd} vertex. It represents the execution order of the command inside the sequence.
\end{itemize}

Let us remember that in our system architecture, the recommendations make sense only under the same scope. For such a reason, the IDs of some vertices are based on scopes (for example the IDs of \textit{cmd} vertices). As a result, a command executed under multiple scopes is represented by a different vertex for each scope, since each of them represents a different entity. This allows to preserve additional relations, such as the number of times each user executed such command under the different scopes. Additionally, the vertex ID of \textit{file} vertices is also based on the paths of the files, since files sharing the same name but located in different paths are different entities.

After its creation, the SRE behaviour knowledge graph is then populated with the processed data and the relationships learned during the SRE behaviour knowledge extraction process. Its content can be updated, for example weekly or monthly, when new shell data is available, after the execution of the SRE behaviour knowledge extraction on such new data.

\section{Knowledge Graph Based Recommender System}
In this section, we describe our recommender system to provide personalized recommendations for each recommendation request by employing the knowledge graph we designed in the previous section. The recommender system is composed by command recommender and sequence recommender.

\subsection{Command Recommender}
\label{sec:com_rec}
In this section, we describe how \alg\ employs the knowledge graph to recommend commands. The idea is to recommend commands that continue part of a command a user is typing in real-time. Thus, given the portion $p$ of a command that the user $u$ is typing inside a session associated with the IP $i$ and the scope $s$, the idea is to first retrieve from the knowledge graph commands that continue the portion $p$ and that have been executed under scope $s$ in the shell data. Then, such candidate commands are ranked considering a scoring function based on their similarity with $p$, and on how many times they have been executed under scope $s$, by user $u$, and inside sessions associated with IP $i$ in the shell data.
 \begin{figure*}[]
\centerline{\includegraphics[width=0.9\linewidth]{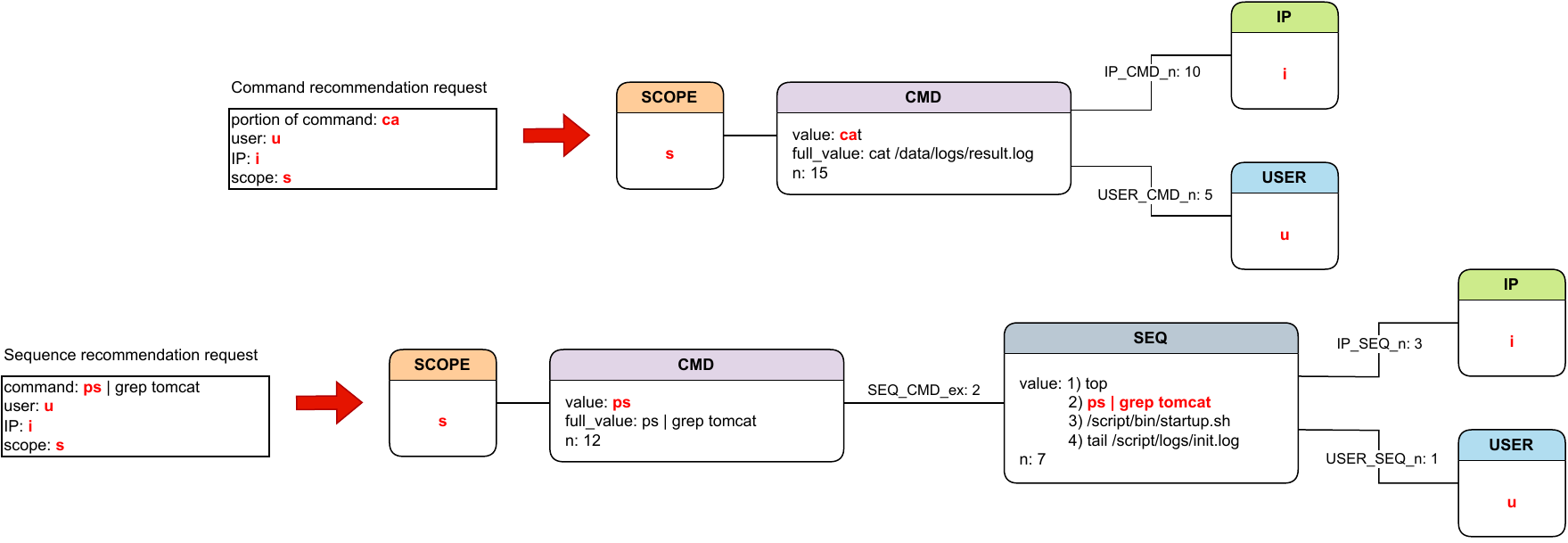}}
	\caption{Example of command recommendation (up) and sequence recommendation (down).}
	\label{fig:cmdRec}
\end{figure*}
 
Fig.~\ref{fig:cmdRec} (up) shows the portion of the knowledge graph accessed to retrieve candidate command recommendations. Starting from the \textit{scope} vertex representing $s$, it is possible to reach all the commands executed under it, represented by the \textit{cmd} vertices connected to it. Let us note that a user, by typing the portion $p$, aims to execute a shell command, which starts with `\textit{cat}', `\textit{ps}', etc., or a custom script, which directly starts with the path of the script. 
In the first scenario, we first tokenize $p$ considering the space as delimiter. The first token is thus a shell command or the prefix of one of them (whether $p$ is composed by a single token). In the knowledge graph, we then only consider \textit{cmd} vertices whose property \textit{cmd.value} starts or is equal to the first token of $p$. In the second scenario, instead, we only consider \textit{cmd} vertices whose property \textit{cmd.value} = `\textit{execute}' (see section~\ref{sec:kg}). Thus, we only consider commands that can continue $p$, obtaining the command candidate set. Finally, for each command $c_r$ in the candidate set, we retrieve \textit{cmd.full\_value}, \textit{cmd.n}, \textit{user\_cmd\_n}, which represents how many times $c_r$ has been executed by user $u$, and \textit{IP\_cmd\_n}, which represents how many times $c_r$ has been executed inside session associated with IP $i$. Let us denote with $\mathcal{C}$ the set of the retrieved candidate commands and their respective information. For each retrieved command $c_r \in \mathcal{C}$, we then compute a score that represents its likelihood to continue the portion $p$ as
\begin{equation*}
\small
	\label{eq:scorecmd}
		\begin{split}
	\texttt{score}(c_r) = \; &\omega_{cmd} \cdot \texttt{sim}(p, c_r) +
	\omega_{user} \cdot \texttt{freq}(u, c_r) \; +\\
	&\omega_{IP} \cdot \texttt{freq}(i, c_r) +
	\omega_{freq} \cdot \texttt{freq}(c_r),
	\end{split}
 \normalsize
\end{equation*}
where:
\begin{itemize}
\item $\omega_{cmd}$, $\omega_{user}$, $\omega_{IP}$, and $\omega_{freq} \in [0,1]$ are 4 weights that define the importance of each component, with $\omega_{cmd} + \omega_{user} + \omega_{IP} + \omega_{freq} = 1$;
\item $\texttt{sim}(p, c_r)$: considers the similarity between $c_r$ and $p$; 
\item $\texttt{freq}(u, c_r)$: considers the number of executions of $c_r$ by user $u$ (\textit{user\_cmd\_n} normalized);
\item $\texttt{freq}(i, c_r)$: considers the number of executions of $c_r$ in sessions associated with IP $i$ (\textit{IP\_cmd\_n} normalized);
\item $\texttt{freq}(c_r)$: considers the number of executions of $c_r$ under scope $s$ (\textit{cmd.n} normalized).
\end{itemize}
To normalize \textit{user\_cmd\_n}, \textit{IP\_cmd\_n}, and \textit{cmd.n}, we divided them by the maximum over their respective values retrieved from the knowledge graph. 

To compute the similarity $\texttt{sim}(p, c_r)$ between the portion $p$ and the whole command $c_r$, any string similarity measure can be employed. In this work, we considered a measure based on the dice coefficient originally proposed to gauge the similarity of two samples. In particular, we first split $p$ and $c_r$ into their character pairs. Then, their similarity is twice the number of character pairs they share divided by the sum of the number of their character pairs. Finally, we normalize all the similarities dividing them by the maximum computed value. We found that this metric performs better than other similarity measures, for example the edit distance. In particular, it allows to better represent similarities between commands that just share a part and this is very advantageous in our scenario. For example, if a user types a shell commands associated with files (e.g., `\textit{cat\textit}', `\textit{tail\textit}') directly followed by a file name or by an extension, we are able to retrieve commands containing such a file or having such an extension.
After computing such a score for each $c_r \in \mathcal{C}$, we then sort the commands in descending order considering the score, and output the top $N$ results, with $N$ the number of commands to recommend to the user.

Let us not that by computing the similarity between commands as described above, we are able to correct possible typos that a user may introduce in $p$, since we are not considering exact matches. The problem remains instead when the user want to execute a shell command and introduces typos in the first token of $p$, i.e., the part of $p$ used to retrieve candidate commands by considering an exact match with \textit{cmd.value}. To avoid such an issue, we designed a simple and effective typo correction system. In particular, before accessing the knowledge graph, we check whether the first space separated token of $p$ is a shell command or a prefix of one of them. (To perform such an operation, it is possible to maintain in memory all the \textit{cmd.value} of all the \textit{cmd} vertices.) If yes, we continue with the classical strategy, otherwise we compute the edit distance between the first space separated token of $p$ and all the \textit{cmd.value}, and consider the most similar one to match \textit{cmd.value} in the knowledge graph.

Finally, let us note that to provide recommendations in real-time while the user is typing, efficiency is of key importance. Thus, we designed a caching system to increase the efficiency by avoiding useless retrieval and re-computation. In particular, when we return command recommendations, we store all the information retrieved from the knowledge graph and the components of $\texttt{score}(c_r)$. Whether the system receives a new request for command recommendations, and the information that would be used in the retrieval phase for this new request (i.e., \textit{cmd.value}, scope $s$, user $u$, and IP $i$) are the ones stored from the previous request, we directly use the old information, without accessing the knowledge graph. Thus, to rank the candidate commands, we just need to compute $\texttt{sim}(p, c_r)$, since $p$ is the only component that changed, and use the other components of the scoring function stored in the caching system to compute $\texttt{score}(c_r)$.

To conclude, let us note that the 4 weights in $\texttt{score}(c_r)$ can be automatically optimized considering a feedback system after that \alg\ is deployed, with the aim of showing the recommendations that the users accept in higher positions.  

\subsection{Sequence Recommender}

In this section, we describe how \alg\ employs the knowledge graph to recommend sequences. The idea is to recommend sequences that continue the operation that a user is executing. Thus, given the command $c$ that the user $u$ has just executed inside a session associated with the IP $i$ and the scope $s$, the idea is to first retrieve from the knowledge graph sequences that continue commands similar to $c$ and that have been executed under scope $s$ in the shell data. Then, such candidate sequences are ranked considering a scoring function based on their similarity with $c$, and on how many times they have been executed under scope $s$, by user $u$, and inside a session associated with IP $i$ in the shell data.

Fig.~\ref{fig:cmdRec} (down) shows the portion of the knowledge graph accessed to retrieve candidate sequence recommendations. Starting from the \textit{scope} vertex representing $s$, it is possible to reach all the commands executed under it whose properties \textit{cmd.value} are the same of the one of $c$, e.g., the shell command `\textit{cat}', `\textit{ps}', etc., or the keyword `\textit{execute}' (see section~\ref{sec:kg}). Thus, we only consider commands that can represent the operation executed by $c$. We then reach all the sequences that contain such commands, obtaining the sequence candidate set. Finally, for each sequence in the candidate set, we retrieve \textit{seq.value}, \textit{seq.n}, \textit{user\_seq\_n}, which represents how many times the candidate sequence has been executed by user $u$, \textit{seq\_IP\_n}, which represents how many times the candidate sequence has been executed inside session associated with IP $i$, \textit{cmd.full\_value} of the command from which we accessed the candidate sequence, and \textit{seq\_cmd\_ex}, which represents the position of such a command inside the candidate sequence, i.e., its execution order. Let us denote with $\mathcal{S}$ the set of the retrieved candidate sequences and their respective information. For each retrieved sequence $s_r \in \mathcal{S}$, with $c_r$ the command from which we accessed it, we then compute a score that represents the likelihood of $s_r$ to continue the operation that user $u$ is executing as
\begin{equation*}
\small
	\begin{split}
		\texttt{score}(s_r) = \; &\gamma_{cmd} \cdot \texttt{sim}(c, c_r) +
		\gamma_{user} \cdot \texttt{freq}(u, s_r) \; +\\
		&\gamma_{IP} \cdot \texttt{freq}(i, s_r) +
		\gamma_{freq} \cdot \texttt{freq}(s_r),
	\end{split}
 \normalsize
\end{equation*}
where:
\begin{itemize}
	\item $\gamma_{cmd}$, $\gamma_{user}$, $\gamma_{IP}$, and $\gamma_{freq} \in [0,1]$ are 4 weights that define the importance of each component, with $\gamma_{cmd} + \gamma_{user} + \gamma_{IP} + \gamma_{freq} = 1$;
	\item $\texttt{sim}(c, c_r)$: considers the similarity between $c_r$ and the executed command $c$; 
	\item $\texttt{freq}(u, s_r)$: considers the number of executions of $s_r$ by user $u$ (\textit{user\_seq\_n} normalized);
	\item $\texttt{freq}(i, s_r)$: considers the number of executions of $s_r$ in sessions associated with IP $i$ (\textit{IP\_seq\_n} normalized);
	\item $\texttt{freq}(s_r)$: considers the number of executions of $s_r$ under scopes $s$ (\textit{seq.n} normalized).
\end{itemize}
To normalize \textit{user\_seq\_n}, \textit{IP\_seq\_n}, and \textit{seq.n}, we divided them by the maximum over their respective retrieved values.

To compute the similarity $\texttt{sim}(c, c_r)$ between the commands $c$ and $c_r$, any string similarity measure can be employed. In this work, we employed the Jaccard similarity between the two tokenized commands $c$ and $c_r$. In particular, we first tokenize $c$ and $c_r$ considering spaces and the path components, as explained in section~\ref{sec:patAgg}. Then, we compute the Jaccard similarity between the two tokenized commands considering the number of tokens they share. Finally, we normalize all the similarities, dividing them by the maximum computed value.
After computing such a score for each  $s_r \in \mathcal{S}$, we then rank the sequences in descending order considering the score, and output the top $N$ results, with $N$ the number of sequences we want to recommend to the user. Note that for each sequence $s_r$, we output the portion of the sequence that continue $c_r$, i.e., all the commands that are executed after position \textit{seq\_cmd\_ex}. A user can decide to execute all such commands or just a sub-sequence of them.

Let us not that while this strategy can be applied to every type of commands, our knowledge graph and recommender system allow to define ad-hoc strategies to improve the recommendation performance for specific categories of commands. Let us consider, for example, shell commands associated with files, which represents a large portion of IT system operations. Suppose that a user has just executed the command $c =$ `\textit{cat /data/logs/result.log $|$ grep error}' and that one of the sequences in the knowledge graph contains the command `\textit{grep error /data/logs/result.log}'. Since such commands represent the same operation, the sequence containing the later is a good recommendation for $c$, but such a sequence would not be in the candidate set considering the standard strategy since the type of the two commands (`\textit{cat}' and `\textit{grep}) are different.
However, by extracting from $c$ the path `\textit{/data/logs/}' and the file `\textit{result.log}' that $c$ accessed, it is possible to consider the \textit{path} vertex representing `\textit{/data/logs/}' and the \textit{file} vertex representing `\textit{result.log}' to reach all the commands, executed under scope $s$, that are associated with such a file to find the commands $c_r$. The remaining retrieval phase and the sequence ranking are then the same of the ones of the standard strategy.

To conclude, let us note that the 4 weights in $\texttt{score}(s_r)$ can be automatically optimized as discussed in section~\ref{sec:com_rec}.

\section{Results and empirical evaluation}
\begin{figure}
        \centering
        \begin{subfigure}[b]{0.49\linewidth}
            \centering
            \includegraphics[width=\linewidth]{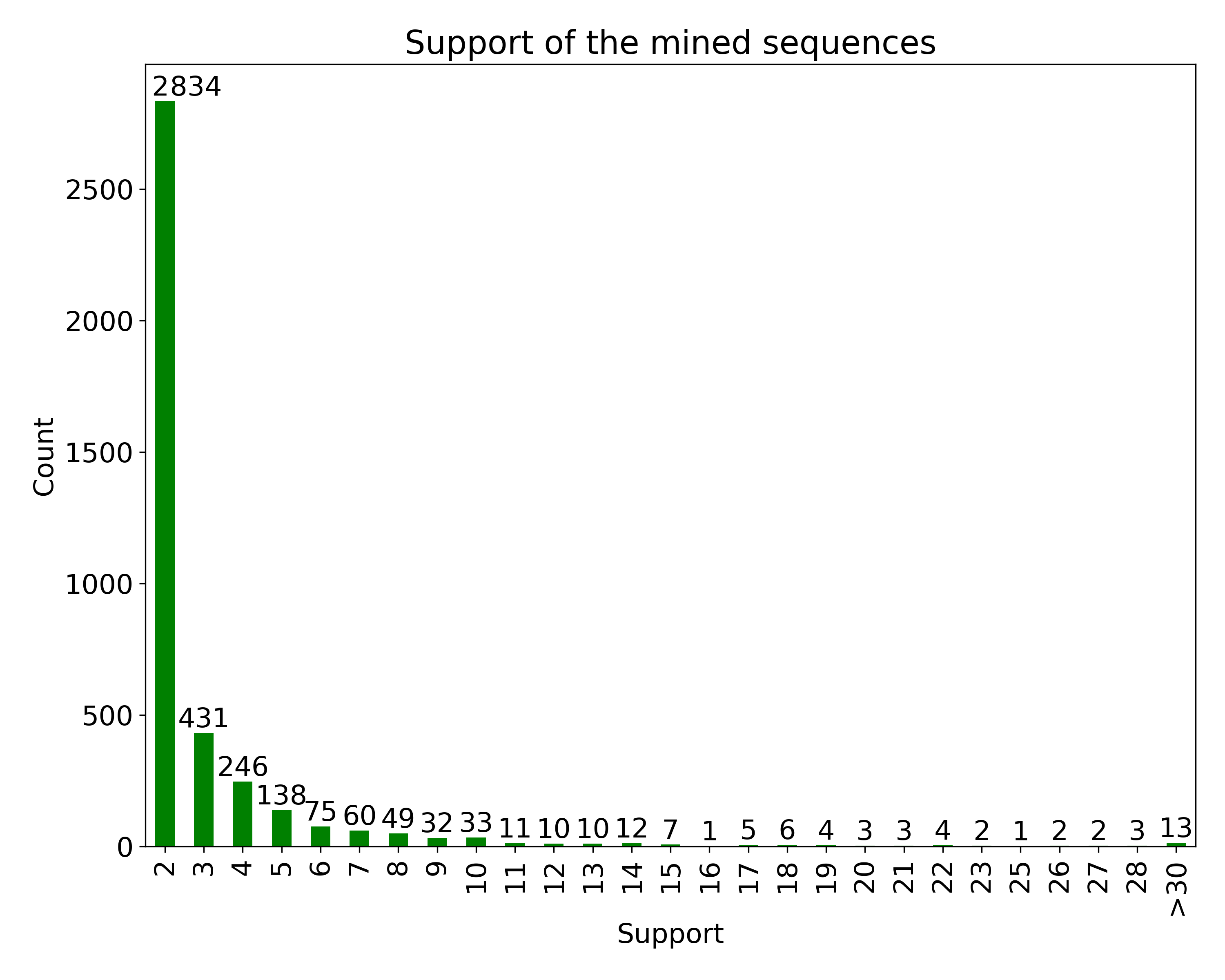}
    \end{subfigure}
        \hfill
        \begin{subfigure}[b]{0.49\linewidth}  
            \centering 
            \includegraphics[width=\linewidth]{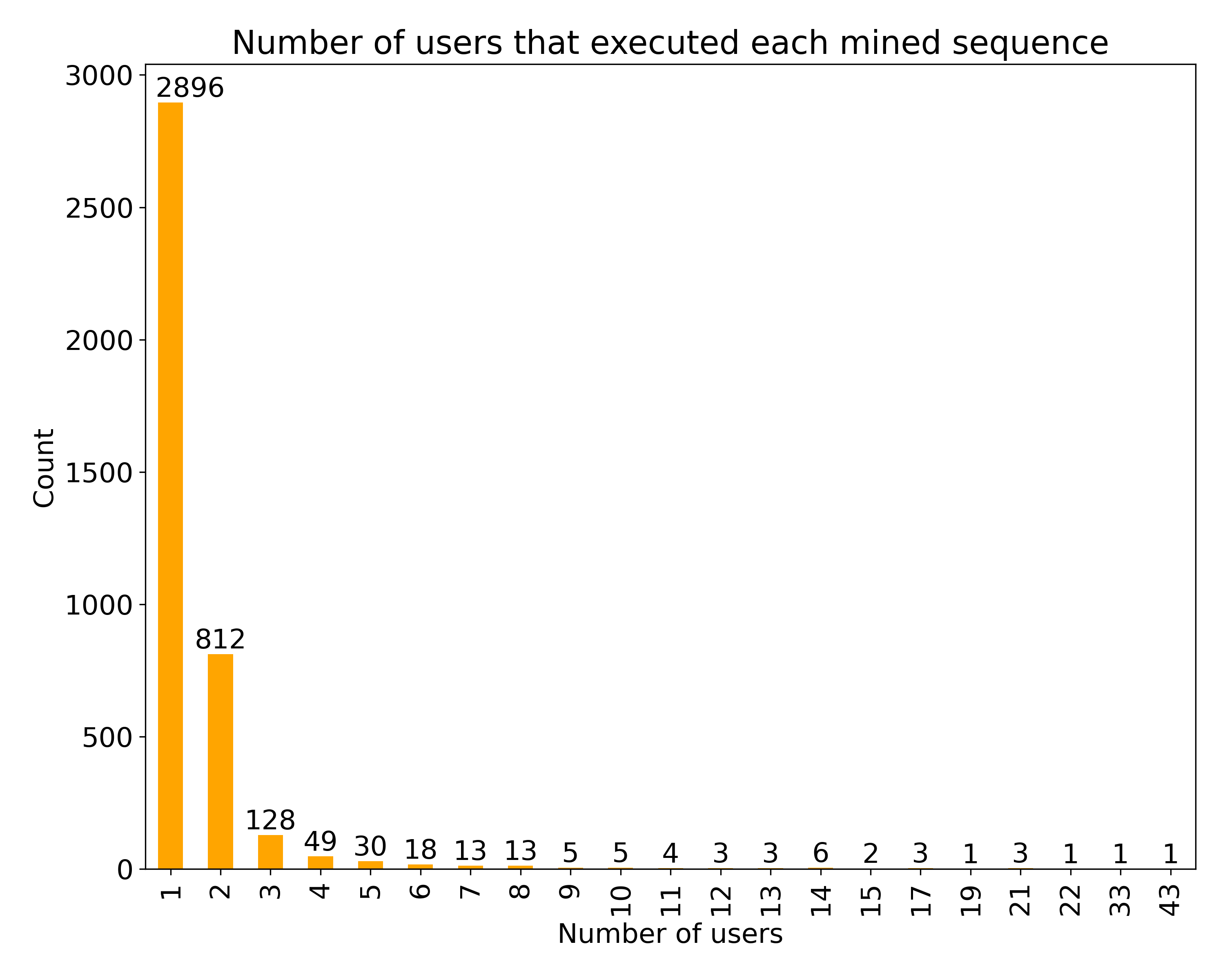}
    \end{subfigure}
    \hfill
        \begin{subfigure}[b]{0.49\linewidth}  
            \centering 
            \includegraphics[width=\linewidth]{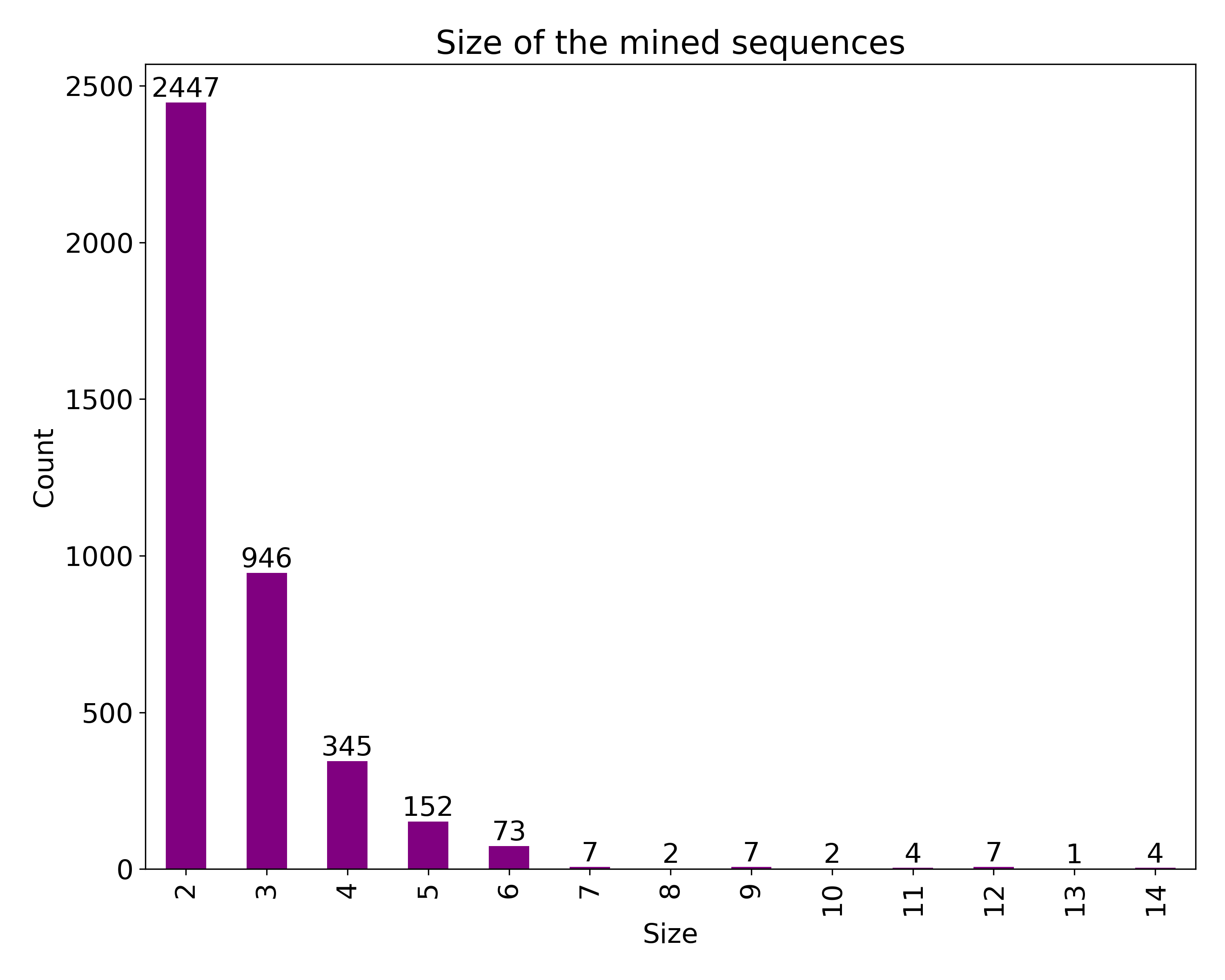}
    \end{subfigure}
    \caption{Results for SRE behaviour pattern mining. It shows the supports of the mined sequences (left), their sizes (right), and the number of users that executed them (down).} 
        \label{fig:min}
    \end{figure}

In this section, we report our results in extracting the SRE behaviour knowledge from real shell data and discuss the estimated efficiency improvements that \alg\ can provide.
\subsection{Data, Implementation Details, and Environment}
We collected the shell commands executed by the SREs of our company in 1 month, obtaining $|\mathcal{D}| = 29859$ sessions executed by 607 users under 58 different scopes. Each session was composed on average by 8.30 commands and had an average duration of almost 3 minutes. The total time spent by SREs in executing shell commands was more than 1418 hours (${\sim}$59 days). Thus, to be able to reduce such a high time by improving the SRE operation efficiency can provide significant business benefits for our company.
We implemented \alg\ in Python 3.10, employing the SPAM~\cite{ayres2002sequential} implementation provided by the SPMF library~\cite{fournier2016spmf} to mine sequential patterns and Nebula graph 3.2.0\footnote{\url{https://www.nebula-graph.io/}} to implement the knowledge graph. Our recommender system is implemented as a Flask service connected with Nebula graph using Nebula's Python API and considers a request-response system to dialogue with a chatOps application internal of our company that allows to execute shell commands and to show the received recommendations. A prototype of \alg\ has been deployed to a server with 64 GB of RAM and an Intel i7-7700K@4.20GHz CPU.

\begin{figure}
        \centering
        \begin{subfigure}[b]{0.49\linewidth}
            \centering
            \includegraphics[width=\linewidth]{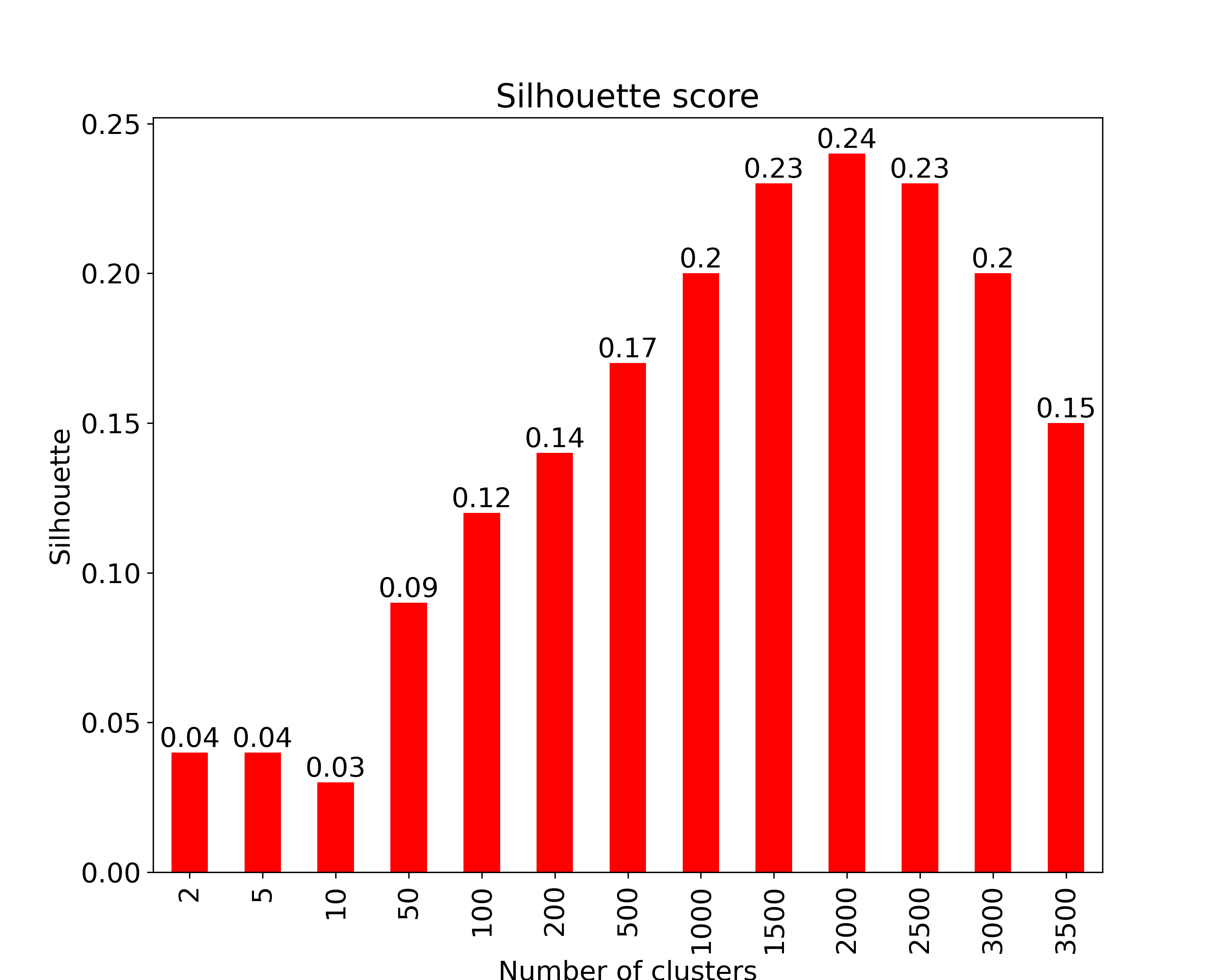}
    \end{subfigure}
        \hfill
        \begin{subfigure}[b]{0.49\linewidth}  
            \centering 
            \includegraphics[width=\linewidth]{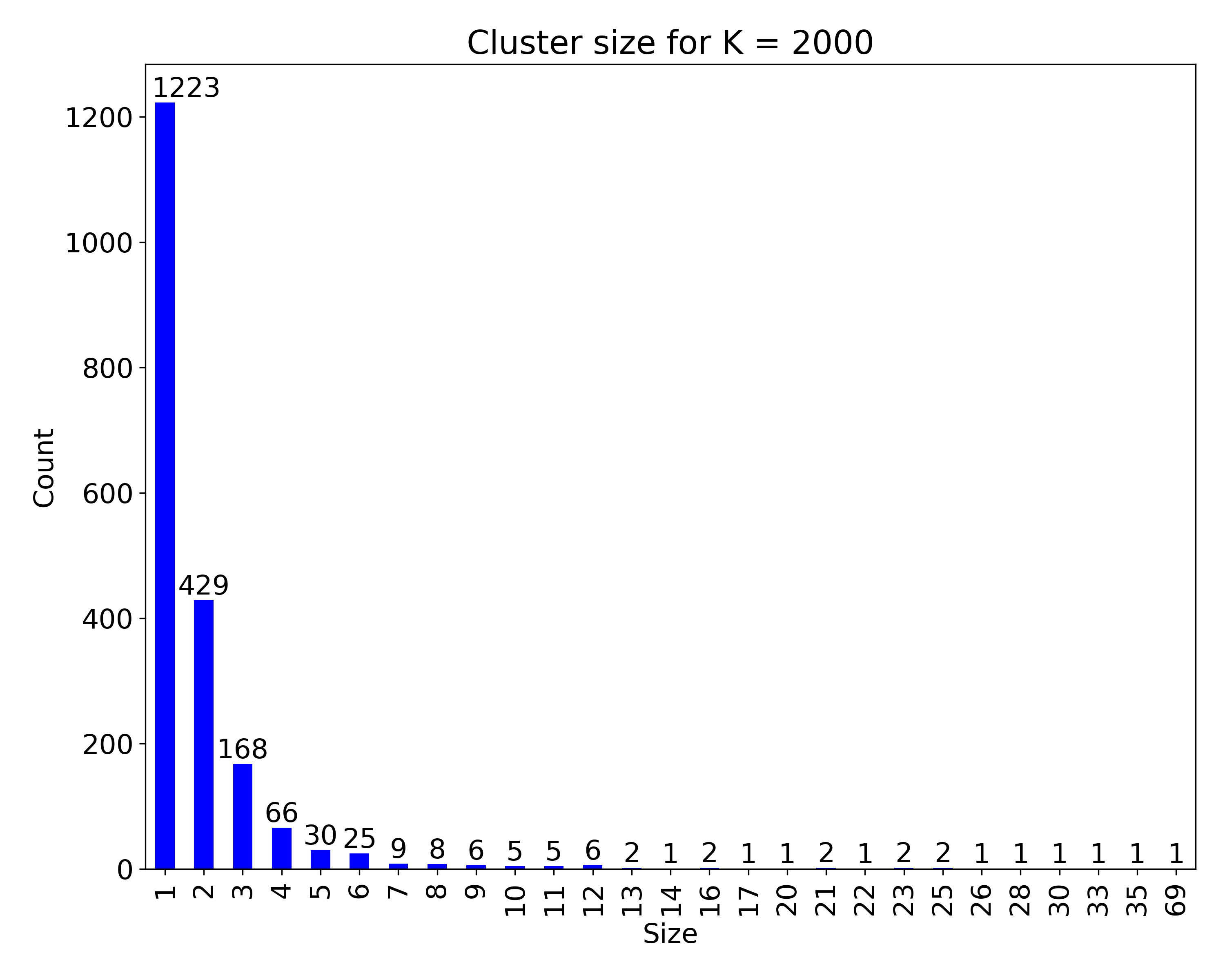}
    \end{subfigure}
    \caption{Results for SRE behaviour pattern aggregation. It shows the silhouette score changing the number of clusters K (left) and the size of the clusters with K = 2000 (right).} 
        \label{fig:clu}
    \end{figure}

        \begin{figure}[]
\centerline{\includegraphics[width=0.9\linewidth]{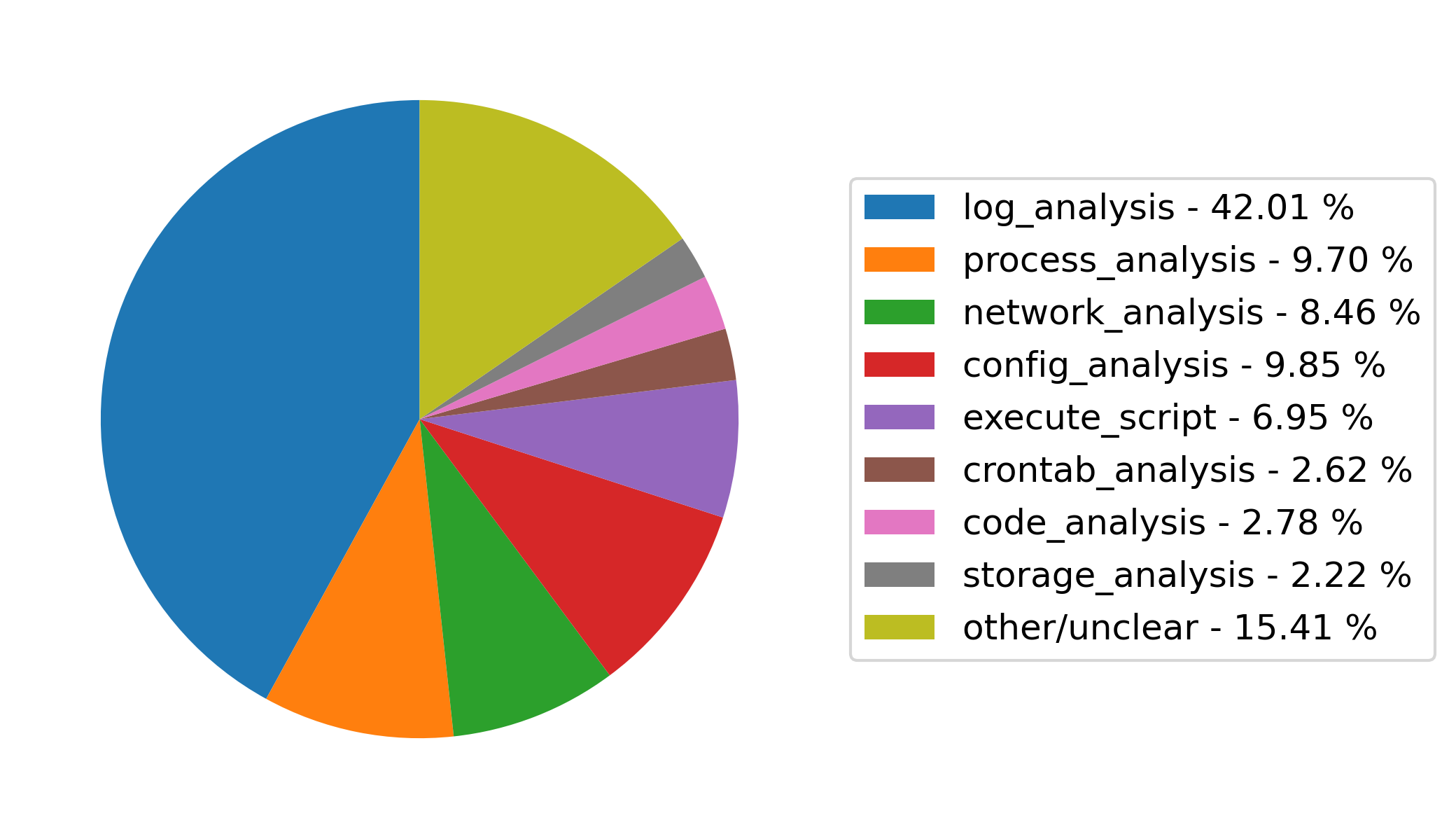}}
\caption{Results of the SRE intent definition. It shows the percentage of classified commands into each intent.}
\label{fig:pie}
\end{figure}

\subsection{SRE Behaviour Knowledge Extraction Results}

In this section, we discuss the results of our approach to extract SRE behaviour knowledge. We considered $min\_supp = 2$ in the parsing and processing phase, obtaining 18524 commands executed by 605 users. Then, we mined all the sequences that appeared in at least 2 sessions ($\theta = 2 / |\mathcal{D}|$), limiting their size from 2 to 20 commands, and considering a maximum gap of 5. Fig.~\ref{fig:min} shows some statistics about the mined sequences. We obtained 3997 sequences, of size ranging from 2 to 14 commands, showing that the mined sequences very often represents complex operations composed by multiple commands that would not be easy to manually define. The mined sequences were executed in a number of sessions, i.e., have a support, ranging from 2 to 101, showing that the sequences are execute many times and thus represents operations that is useful to recommend over time. In addition, they are often executed by more than one user, up to a maximum of 43, making them eligible to be shared to all the SREs.    
By showing the mined sequences to the SREs, they recognized almost all the sequences as operations that they usually execute, confirming the validity of our method and highlighting that by mining frequent sequences, we are able to extract complex operations composed by many commands executed multiple times in the past. We then applied our SRE behaviour pattern aggregation method to the mined sequences, obtaining an optimal number of clusters $k = 2000$. Fig.~\ref{fig:clu} shows some statistics about the clusters. The majority of the clusters were composed by a single sequence, but some of them contained more than one sequence (up to 69). By providing the clustered sequences to the SREs, they used them to define 35 macros, confirming that clustered sequences, in some cases, represented similar operations that could be grouped in a single macro. In other cases, instead, to maintain the original sequences was a better choice. 
Finally, we classified the shell commands into SRE intents. Fig.~\ref{fig:pie} shows the results. With our approach, we classified almost $85\%$ of the shell commands, with the majority of them representing operations associated with files. Among unclear commands, we found many commands that accessed files without extensions or with extensions not covered by our rules. These results show that the 8 general SRE intents we defined allow to represent the majority of the executed operations, obtaining more concise and general commands that directly describe the aim of the shell commands.     

\subsection{Estimated Efficiency Improvement and Response Time}

Let us note that to exactly quantify the efficiency improvement that \alg\ can provide on a large scale is not an easy task. Thus, we focused on estimating 3 aspects considering the real shell data. Table~\ref{tab:res} reports the estimated improvements.

The first aspect aims to quantify the reduced number of command lines the users need to type to execute a command using \alg, by comparing the number of commands in the shell data and the number of commands in the processed data.
Indeed, almost every session in the shell data contained many commands that were not essential to perform the required operations, such as long sequences of `\textit{cd}' commands to change file-system location, erroneous commands, etc. Thus, this metric aims to quantify the reduced number of command lines the users need to type to execute a command, without considering the erroneous and not-useful commands that we observed in real shell data since \alg\ allows to avoid them. Given a sequence of commands $x$ representing a session in the shell data, and given a sequence of commands $y$ representing the same session in the processed data, we estimated the improvement of such a session as
$1 - \left( \frac{|y|}{|x|}\right)$. The average improvement is then the average over all the sessions. 
The estimated average command reduction that \alg\ can allow is 43\%, with a maximum of 99\% for a single session.

The second aspect aims to quantify the reduced number of characters the users need to type to execute a command associated with a file considering the command recommender.
Let us remember, from section~\ref{sec:com_rec}, that the string similarity measure we employed allows to better represent similarities between commands that just share a portion. Thus, by typing a shell command associated with files (e.g., `\textit{cat}', `\textit{tail}', etc.) directly followed by a file name (e.g., `\textit{cat result.log}'), it is possible to retrieve the whole commands accessing such a file (e.g., `\textit{cat /opt/hw/configuration/logs/result.log}'). Note that this is not always true, since, depending on the commands, in some cases more characters (including a portion of the path of the file) or less characters (just a portion of the file name) are necessary. However, we found that, on average, this strategy can provide good estimates of this aspect.    
Thus, for each command associated with a file, the reduced number of characters to type can be estimated as
1 minus the ratio between the number of characters the user needs to type to obtain the recommendation (e.g., $|$`\textit{cat result.log}'$|$ = 14) and the number of characters in the whole shell command (e.g., $|$`\textit{cat /opt/hw/configuration/logs/result.log}'$|$ = 41).
The average improvement is then the average over all the commands associated with files. The estimated average character reduction that our command recommender can allow is 72.5\%. Let us remember that shell commands associated with files represent a large portion of the IT system operations and thus such a reduction can provide significant improvements.
 
The third aspect aims to quantify the reduced number of command lines the users need to type to execute a sequence of commands or a macro considering the sequence recommender.
Given a mined sequence or a macro $s$, we computed its benefits as 
$1 - \left( \frac{|1|}{|s|}\right)$, 
since by typing the first command of the sequence, or the intent of the macro, \alg\ provides as recommendations all the other commands that can be directly executed without typing other commands. The average improvement is then the weighted average of the single sequence/macro improvements considering the sequences' support, i.e., the number of sessions in which they appear, as weight. The estimated average command reduction that our sequence recommender can allow is 57\%, with a minimum and maximum reduction of 50\% and 93\%, respectively, for a single sequence/macro.

Overall, these estimates show that \alg\ can improve the SRE operation efficiency in executing shell commands and that to preserve and to re-utilize SRE shell operation knowledge can provide significant benefits.

\begin{table}[]
\centering
\caption{Estimated efficiency improvements.}
\label{tab:res}
\begin{tabularx}{\linewidth}{X>{\hsize=.5\hsize\raggedleft\arraybackslash}X}
\toprule
\multicolumn{1}{c}{Aspect}                                                      & \multicolumn{1}{c}{Estimated improvement} \\ \midrule
Reduced number of command lines to type for executing a command                    &  $43\%$                                     \\ \midrule
Reduced number of characters to type for executing a command associated with files & $72.5\%$                                   \\ \midrule
Reduced number of command lines to type for executing a sequence of commands        & $57\%$                                    \\ \bottomrule
\end{tabularx}
\end{table}

To conclude, we discuss the time required by \alg\ to provide recommendations.
The command recommender requires a time ranging from a few milliseconds to a maximum of almost $200ms$, which is consistent with real-time requirements. (When the cache system is used, instead, the response time is always less than $10ms$.) The sequence recommender, instead, requires a time ranging from $50ms$ to almost $300ms$. Let us note that while for the command recommender we need to provide recommendations in real-time to show them while the user is typing, for the sequence recommender such a constraint is less important since the sequence recommendations are shown after the user executed a command and observed its results. 

\section{Conclusion}
In this work, we propose \alg, a knowledge graph model to preserve SRE knowledge learned from shell data and to recommend shell commands for IT operations to improve the SRE operation efficiency. In particular, we introduce a method to automatically extract SRE knowledge from shell data and we model a knowledge graph to preserve such a knowledge and its relations. The knowledge graph is then used by our recommender system to recommend (sequences of) shell commands considering the relations between the involved entities learned from the data in order to share such an operational knowledge to all the SREs. Our final discussion highlights the benefits that \alg\ can provide in improving the SRE operation efficiency. To conclude, an interesting future direction is the inclusion of natural language functionalities to execute shell commands employing our SRE behaviour knowledge graph.

\balance
\bibliographystyle{splncs04}
\bibliography{bib}

\end{document}